\documentclass[a4paper,11pt]{article}
\pdfoutput=1 

\usepackage{jcappub} 

\usepackage[T1]{fontenc} 
\usepackage{float}

\usepackage[normalem]{ulem}

\title{\boldmath Not So Minimal Warm Inflation}

\author[a]{Mar Bastero-Gil}
\author[a]{Pedro García Osorio}
\author[b]{António Torres Manso}

\affiliation[a]{Departamento de F\'{i}sica Te\'{o}rica y del Cosmos and CAFPE,\\
	Universidad de Granada, Campus de Fuentenueva, E-18071 Granada, Spain}
\affiliation[b]{CFisUC, Departamento de Física, Universidade de Coimbra,   \\ Rua Larga, 3004-516 Coimbra, Portugal}

\emailAdd{mbg@ugr.es}
\emailAdd{pgosorio@correo.ugr.es}
\emailAdd{atmanso@uc.pt}

\abstract{An axion-like inflaton coupled to non-Abelian gauge bosons provides a compelling microphysical framework for warm inflation. Starting even from cold initial conditions, in these systems, sphaleron heating may generate thermal friction sufficient to sustain finite temperatures throughout the inflationary epoch. Insisting on shift-symmetric potentials, in this work we revisit the viability of these scenarios under the designation of Minimal Warm Inflation. We examine both observational constraints and model-building limitations on models with a hierarchy between the decay constants appearing in the friction rate and in the inflaton potential. We conclude that the popular clockwork mechanism cannot generate the required hierarchy; however, partial-wave unitarity bounds admit effective descriptions that remain consistent with observations.}

\begin{document}
\maketitle
\flushbottom

\section{Introduction}
\label{sec:intro}

In standard cosmology, cosmological inflation, an early phase of accelerated expansion, is the leading framework for resolving the flatness and horizon problems \cite{Guth:1980zm,inflinde,infalbrecht}. This accelerated expansion is generally driven by a scalar field called the inflaton $\phi$, undergoing slow-roll evolution, whose quantum vacuum fluctuations provide a natural mechanism for generating the anisotropies observed in the Cosmic Microwave Background (CMB). A critical challenge, however, is understanding the transition from inflation to the well-established hot big bang cosmology. This includes the reheating epoch, during which the universe becomes radiation-dominated and the conditions are established for Big Bang Nucleosynthesis (BBN), the process responsible for synthesizing light nuclei \cite{Fields:2014uja,ParticleDataGroup:2020ssz}. Achieving this transition requires coupling the inflaton to other particle species \cite{Figueroatanin}. In this regard inflation models are typically classified in two main categories: cold inflation (CI) and warm inflation (WI) models.

In the more conventional CI models there is an assumption of near-zero temperature until the post-inflationary reheating phase that generates the Hot Big Bang content. On the other hand, during WI \cite{Berera:1995ie}, the inflaton field interacts continuously with a thermal radiation bath during the accelerated expansion phase, generating entropy (particles) through dissipative dynamics. This construction provides both a mechanism to sustain slow-roll evolution with dissipative dynamics and avoids the need for a separate reheating phase \cite{Berera:1996nv,Berera:1996fm,Berera:1998gx}. Arising from a dominant contribution of thermal fluctuations, as opposed to CI quantum fluctuations, WI provides distinct observational signatures, such as modified primordial power spectra and non-Gaussianities, testable against cosmic microwave background (CMB) data \cite{Spectrum3,Bastero-Gil:2014raa,Benetti:2016jhf,Bastero-Gil:2017wwl, Arya:2017zlb,Kumar:2024hju}. 
In order to ensure the flatness of the inflaton potential and prevent it from receiving large thermal corrections which could disrupt the inflationary attractor solution, diverse particle physics models with integrated symmetries have been built. While early models explored supersymmetric properties, more recent analyses have turned to constructions where the inflaton would be a pseudo-Goldstone boson \cite{Bastero-Gil:2016qru,Bastero-Gil:2019gao} or would have axion-like couplings to gauge fields \cite{Mishra:2011vh,Visinelli:2011jy,Kamali:2019ppi,Berghaus_2020,DeRocco:2021rzv}. See \cite{Kamali:2023lzq} for a more complete review of recent developments.

The models we explore in this paper, the Minimal Warm Inflation (MWI) scenarios, consider the last example where the inflaton is the axion of a non-Abelian gauge sector. In these models, starting from a zero temperature initial condition, the non-perturbative sphaleron transitions lead to the generation of a thermal plasma \cite{Kuzmin:1985mm, McLerran:1990de, Berghaus_2020, Laine_2021, Klose:2022rxh, Mirbabayi:2022cbt, Kolesova:2023yfp, Berghaus:2024zfg}, the so called sphaleron heating, while at the same time an axial symmetry protects the potential from thermal corrections. Within this framework there has been a proposal of a WI construction with a minimal extension of the standard model (SM) of particle physics\footnote{Other attempts of WI motivated models with SM interactions can be found in \cite{Dymnikova:2000gnk,Dymnikova:2001jy,Kamada:2009hy,Levy:2020zfo}.}, where the inflaton is coupled to gluons via the standard axial interaction \cite{Berghaus:2025dqi}. The validity of the model was subsequently analyzed with numerical tools in \cite{ORamos:2025uqs}. { Moreover, the contributions of an effective chemical potential generated by a thermal plasma composed purely of gauge bosons have also been considered before \cite{Broadberry:2025ggb}, and these may counteract the effect of the pseudoscalar coupling, in turn decreasing the friction effect. Nevertheless, these effects have only been qualitatively estimated when taking into account the important non-perturbative evolution for the non-Abelian dynamics. Therefore, we will not consider them in our quantitative analysis.}

To support the theoretical motivations of MWI, a non-perturbatively generated potential respecting the discrete shift symmetry of the axion is highly desirable. However, difficulties have already been observed in reconciling warm axion inflation with standard properties of an axion in numerical studies \cite{Klose:2022rxh}. Furthermore, it has been theoretically demonstrated that "demanding friction to be dominated by sphaleron heating together with the requirement of preserving the usual properties of a “vanilla” axion in fact excludes slow-roll inflation" \cite{Zell:2024vfn}. 

In this paper we maintain the effort of reconciling shift symmetric potentials with the dynamics of MWI\footnote{See \cite{Ito:2025lcg} for an analysis on multi-natural inflation models.}. We will study the observationally viable scenarios where the axion gauge field coupling is parametrically large relative to the field range of the axion, meaning, two different effective decay constants in the potential and in the friction coefficient. This allows a two scale system, different from the "vanilla" axion studied in \cite{Zell:2024vfn}, that makes the periodic potential compatible { with observations for a relatively large hierarchy between the scales. We will attempt then to embed the phenomenological viable models within UV-complete descriptions}. Based on the analysis developed in \cite{Agrawal:2018mkd,Bagherian:2022mau}, we will see that the commonly used clockwork constructions cannot generate the required hierarchy, but  EFT partial-wave unitarity bounds may allow some of the phenomenologically viable models. In the last part of this paper we will highlight the relevance of a clear physical definition for the inflaton's fluctuation oscillation frequency in the case of harmonic potentials. 

This paper is organized as follows: In section \ref{sec:section2} we review the standard settings for Warm Inflation, and in particular for MWI, in order to study its observational predictions. In section \ref{Sec:parameterspace}, we analyze the phenomenological implications of introducing a hierarchy between the scales in the inflaton potential and the dissipative coefficient, whereas in section \ref{Sec:Modelbulding} we study the viability of such a hierarchy in the model-building sense. Next, in section \ref{sec:Disc} we review some of the initial assumptions related to the initial temperature of the system and the frequency of the inflaton's fluctuations, and finally in \ref{sec:Concl} we present our conclusions and proposals for future work.

\newpage

\section{Minimal warm inflation with medium response}
\label{sec:section2}

In WI, the equations of motion for the background dynamics describe the evolution of the classical inflaton field, $\phi$, and the radiation bath energy density, $\rho_R$, given by \cite{Bastero-Gil:2009sdq}:
\begin{align}
&\ddot{\phi}+(3H+\Upsilon)\dot{\phi}+\partial_\phi V = 0,\label{eq:EoM_Phi}\\
 &   \dot \rho _R + 4H\rho_R = \Upsilon\dot\phi^2,\label{eq:EoM_Rad} 
\end{align}
where $V(\phi)$ is the inflaton potential, $H$ the Hubble scale and $\Upsilon$ the dissipative coefficient that arises from the interactions between the inflaton and the radiation degrees of freedom. The Hubble scale can be expressed in terms of the total energy density of the universe,
\begin{equation}
    H^2=\frac{1}{3m_p^2}\left(\frac{1}{2}\dot\phi^2 + V(\phi)+\rho_R\right),
    \label{eq:EoM_Hubble}
\end{equation}
and the temperature $T$ of a thermalized radiation bath can be obtained as
\begin{equation}
\rho_R=g_*\frac{\pi^2}{30}T^4,
\end{equation}
where $g_*$ is the effective number of radiation degrees of freedom. In this work we will assume that radiation is composed by the gauge bosons of a $SU(N_c)$ gauge group together with the thermalized inflaton\footnote{A discussion on the consequences of a thermalized (or not thermal) inflaton is presented in Appendix \ref{Sec:thermalinf}.}, resulting in $g_*=2(N_c^2-1)+1$.

The slow-roll parameters in WI can be expressed in terms of the standard CI slow-roll parameters $\epsilon_V$ and $\eta_V$ \cite{Bastero-Gil:2009sdq},
\begin{flalign}
   & \epsilon_H=-\frac{\dot{H}}{H^2}\simeq\frac{\epsilon_V}{1+Q}=\frac{m_p^2}{2(1+Q)}\left(\frac{V'(\phi)}{V(\phi)}\right)^2,\\
    &\eta=\frac{\eta_V}{1+Q}=\frac{m_p^2}{1+Q}\frac{V''(\phi)}{V(\phi)},
\end{flalign}
where $Q\equiv\Upsilon/3H$ is known as the dissipative ratio in WI. This ratio is used to measure the strength of thermal dissipation at a given point in time, differentiating between two regimes in WI: the Weak Dissipative Regime (WDR) { when $Q <1$, and the Strong Dissipative Regime (SDR) when $Q > 1$}.

In the microscopic model that we are studying, MWI, the inflaton is assumed to act like a pseudo Nambu-Goldstone boson { of a certain U(1) Peccei-Quinn-like symmetry. It will be coupled to the corresponding set of non-abelian gauge bosons of a $SU(N_c)$ gauge group, where $N_c>1$ is the number of colors, through an axion-like operator with a certain gauge coupling $g$ and a decay constant $f_a$,}
\begin{equation}
    \mathcal L\supset -\frac{g^2}{64\pi^2}\frac{\phi}{f_a} \varepsilon^{\mu\nu\rho\lambda}F_{\mu\nu}^a F_{\rho\lambda}^a.
    \label{eq:lagrangian}
\end{equation}
The potential energy of the inflaton { is generated by $SU(N_c)$ instantons and takes the following expression},
\begin{equation}
    V(\phi)=\Lambda^4 \left[1-\cos{\left(\frac{\phi}{f_b}\right)}\right],\label{eq:potential1}
\end{equation}
with a single minimum at $\phi=0$ and a shift symmetry $\phi\rightarrow \phi+2\pi f_b$ \cite{Freese:1990rb}. In principle, the scale $f_b$ in the potential is expected to be the same as the scale $f_a$ in the interaction term, but as we will see later this results in predictions that are incompatible with cosmological observations, and therefore it will be convenient to introduce a hierarchy between the two. However, in the initial minimal setting, we will simply assume $f_b=f_a$.

As the inflaton passes part of its energy onto the gauge bosons, these will then quickly thermalize, giving rise to a thermal bath of radiation during the inflationary universe. Given the interaction in (\ref{eq:lagrangian}), the dissipative coefficient in the inflaton's equation of motion will take the following form \cite{Laine_2021,Klose:2022rxh},
\begin{equation}
    \Upsilon=\frac{\alpha ^2 d_A }{f_a^2}\left(
    \kappa  (\alpha  N_c T)^3
    \frac{1+\frac{\omega ^2}{(c_{IR} \alpha ^2 N_c^2 T)^2}}{1+\frac{\omega ^2}{(c_M \alpha N_c T)^2}}
    +
     \coth{\left(\frac{\omega}{4T}\right)}
    \frac{\pi  \omega ^3 }{(4 \pi )^4}
    \right),\label{eq:disip_coef}
\end{equation}
where $\kappa=1.5$, $c_{IR}=106$ and $c_M=5.1$ are constants, $d_A=N_c^2-1$ is the number of gauge bosons in the theory, $\alpha=g^2/4\pi$ where $g$ is the gauge coupling, and $\omega$ is the frequency scale of the inflaton's fluctuations. This scale is usually taken to be the mass of the inflaton when one is near the minimum of the potential, but its value away from the minimum is still unclear, and is often considered to be zero when the second derivative of the potential is negative \cite{Klose:2022rxh}.

The most interesting feature of the dissipative coefficient in equation (\ref{eq:disip_coef}) resides in its non-trivial vacuum structure, since $\omega>0$ results in non-zero dissipation even without an initial radiation bath, unlike the dissipative coefficients that arise from the interaction with fermions and scalars, which are proportional to a power of the temperature \cite{Bastero-Gil:2010dgy}. Theoretically, this could provide a mechanism where the radiation bath can be generated from an initial state with zero temperature as long as $\omega\neq0$.

 In \cite{Klose:2022rxh} it has been argued that the choice of $\omega=m$ or $\omega^2=\max(0,\partial_\phi^2V)$ does not produce significantly different results. To ensure that a thermal bath can be generated before CMB scales, we take $\omega=m$ and we will later revise these assumptions in section \ref{sec:Disc}.

The choice of the number of colors $N_c$ in the gauge group is also relevant for this dissipative coefficient. In this work we have used $N_c=3$ for numerical purposes due to analogy with the strong gauge group as well to facilitate comparison with previous works.

\subsection{Primordial spectrum in WI}
\label{sec:WICMB}

In the presence of a thermal bath, the expression of the scalar amplitude of the primordial spectrum takes the following form \cite{Bastero-Gil:2018uep,Spectrum1,Spectrum2,Spectrum3}:
\begin{equation}
    P_{\mathcal R} = \left(\frac{H_*}{2\pi}\right)^2\left(\frac{H_*}{\dot \phi_*}\right)^2\left(
    1+2n_* + \frac{2\sqrt 3 \pi Q_*}{\sqrt{3+4\pi Q_*}}\frac{T_*}{H_*}\right) G(Q_*),\label{eq:PRWarm}
\end{equation}
where the subscript ``$*$'' indicates the value of the quantities at horizon crossing, and $Q= \Upsilon/3H$ is the dissipative ratio defined above. Here, $n_*$ represents the phase space distribution of the inflaton's fluctuations. If these fluctuations are assumed to have thermalized with the radiation bath (and are therefore thermal fluctuations), the expression for $n_*$ will be a Bose-Einstein distribution at temperature $T_*$, $n_*\approx(e^{H_*/T_*}-1)^{-1}$. 
Although this is not necessarily the case, motivated by \cite{Ferreira:2017lnd,Ferreira:2017wlx,ORamos:2025uqs,Bastero-Gil:2017yzb} we will assume that the inflaton follows a thermal distribution. Furthermore, in Appendix \ref{Sec:thermalinf} we verify that a system where the inflaton does not thermalize cannot satisfy both the observational and the theory constraints discussed later in Section \ref{Sec:Modelbulding}.

The $G(Q_*)$ factor is known as the {\em growing mode} function, which accounts for the fact that when $Q_*\gtrsim1$ the inflaton's fluctuations and the radiation energy density fluctuations become strongly coupled to each other. In this regime the influence of the radiation fluctuations onto the inflation causes the curvature perturbation to grow before horizon crossing. The explicit value of this effect must be computed numerically. The expression for this function depends on the type of dissipative coefficient used and its dependence with respect to temperature. In this work, we use a growing mode obtained in \cite{Benetti:2016jhf,Kamali:2023lzq} for a dissipative coefficient that has a cubic dependence with temperature, $\Upsilon\sim T^3$, since the one in equation (\ref{eq:disip_coef}) behaves as such as soon as $T\gg\omega$, which is largely the case in the strong dissipative regime,
\begin{equation}
    G(Q)=1+4.981 Q^{1.946}+0.127 Q^{4.33}   +1.67\cdot 10^{-5} Q^{6.7837}.
\end{equation}

We also use the expression for $P_{\mathcal R}$ in (\ref{eq:PRWarm}) to compute the scalar spectral index $n_s$ and the running of the spectral index  $\alpha_s$,
\begin{equation}
    n_s-1=\frac{d\ln P_{\mathcal R}}{d\ln k};\,\,\,\,\,\alpha_s=\frac{d n_s}{d\ln k}=\frac{d^2\ln P_{\mathcal R}}{d(\ln k)^2},
\end{equation}
as well as the tensor-to-scalar ratio $r$,
\begin{equation}
    r= \frac{2}{\pi^2 P_{\mathcal R}}\frac{H_*^2}{m_p^2}.
    \label{eq:deltat}
\end{equation}
An analytical expression for $n_s$ in WI can be found in appendix \ref{sec:ns-warm}, while numerical derivations have been used to calculate $\alpha_s$.

The point in time at which horizon crossing takes place depends on the number of e-folds between horizon crossing and the end of inflation, which we will call $N_*$ and is generally assumed to be of order $\mathcal O (50-60)$. Because the exact value depends on reheating parameters \cite{Adshead:2010mc}, we will simply consider the limiting cases $N_*=50$ and $N_*=60$ in the following sections.

\subsection{Observational predictions of MWI}

To test the viability of this setup, we calculate the values of $P_{\mathcal R}$, $n_s$ and $r$ for the potential (\ref{eq:potential1})  and the dissipative coefficient (\ref{eq:disip_coef}). Since the amplitude of the potential $\Lambda$ can be used to fit $P_{\mathcal R}$ to its observational value, we only need to search for combinations of the free parameters $\alpha$ and $f_a=f_b$ that satisfy the observational constraints on $n_s$ and $r$.

It has been shown before that this setup is incapable of maintaining warm inflation in the strong dissipative regime ($Q>1$) for the necessary period of $\mathcal O(50-60)$ e-folds of inflation \cite{Zell:2024vfn}. While this number of e-folds is achievable in the weak dissipative regime ($Q<1$), it requires a scale of $f_b\gtrsim m_p$ in the potential for inflation to be viable.

\begin{figure}[t]
    \centering
    \includegraphics{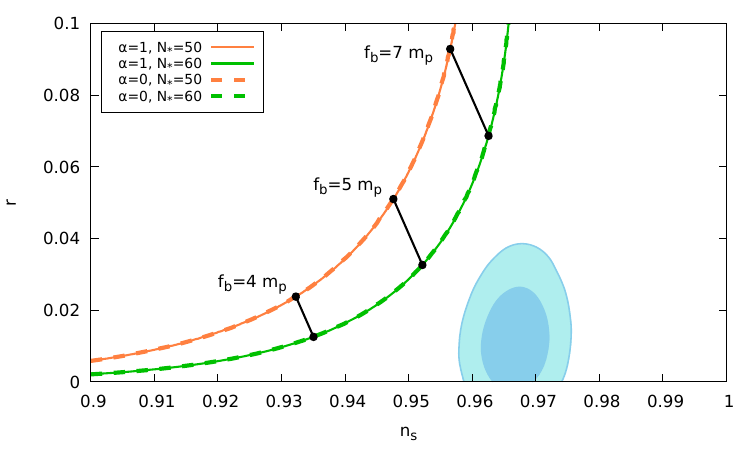}
    \caption{Plot of the scalar spectral index $n_s$ versus the tensor-to-scalar ratio $r$, obtained by varying the parameter $f_b$ (setting $f_a=f_b$) with some values of $f_b$ highlighted in the curve, together with the observational range for $n_s$ and $r$ from the BK18 datasets \cite{BK18}. We compare the results for $N_*=50$ and $N_*=60$, as well as the minimum and maximum values of the parameter $\alpha$.}
    \label{fig:fafb_results}
\end{figure}

Assuming the minimal setup with $f_a=f_b$, we find that simultaneously requiring $\alpha\leq1$ and $f_a\geq m_p$ in the dissipative coefficient yields a $Q_*$ of order $10^{-30}$ or smaller at horizon crossing. In this regime, all dissipative effects become negligible, and observational predictions are left the same as they are in CI, as can be seen in Fig.\ \ref{fig:fafb_results}. The case with maximum thermal dissipation allowed by perturbativity bounds ($\alpha=1$) is indistinguishable from the CI case ($\alpha=0$), showing that, in MWI scenarios, assuming $f_a=f_b$ causes thermal dissipation to have no visible impact on observational predictions. It is then possible to conclude that MWI (assuming that $f_a=f_b$) is excluded by cosmological observations\footnote{Throughout this work we have used Planck-BK18 constraints \cite{BK18}. Although the inclusion of more recent ACT bounds, $n_s=0.974\pm0.003 $ and $r<0.036$, would slightly modify the quantitative limits on some model parameters, it would not modify our analysis and conclusions.}  \cite{BK18}. We will consider the values of $n_s=0.9649\pm0.0042 $ and $r<0.036$.



\section{A phenomenologically viable model with \texorpdfstring{$f_a\ne f_b$}{fa=/=fb}}\label{Sec:parameterspace}

As we have seen in the previous section, on one hand requiring the axion decay constant $f_a$ in the dissipative coefficient to be above the Planck scale results in negligible dissipation. On the other, inflation is not viable unless the scale $f_b$ in the potential is larger than the Planck mass. 
A possible solution for this is to assume an underlying mechanism, causing the scale in the potential $f_b$ to be an effective scale, generating a hierarchy such that $f_a\lesssim m_p \lesssim f_b$. 

\begin{figure}[ht]
    \centering
    \includegraphics{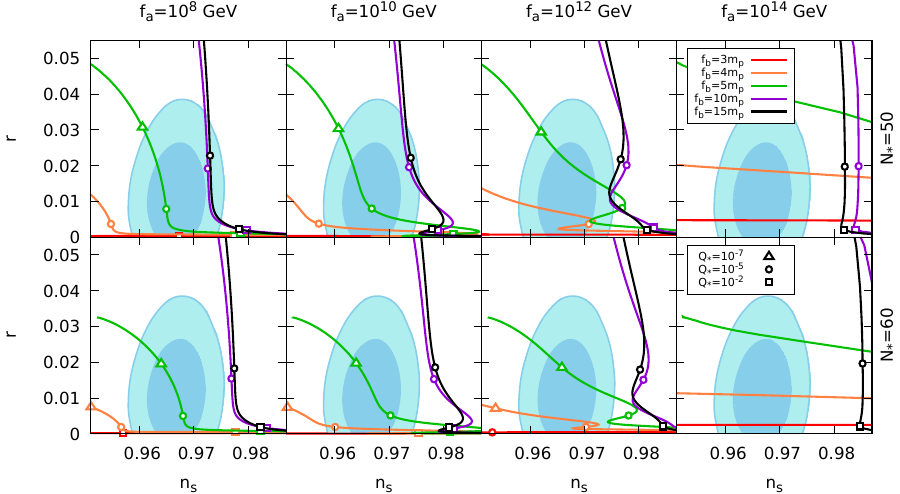}
    \caption{Plot of $n_s$ versus $r$ for different values of $f_a$ and $f_b$, as well as both $N_*=50$ and $N_*=60$. The curves are obtained by varying the parameter $\alpha$. The blue area marks the BK18 observational constraints on $n_s$ and $r$ \cite{BK18}. In each case we have highlighted the points where the dissipative ratio at horizon crossing, $Q_*$, reaches a value of $Q_*=10^{-7}$, $Q_*=10^{-5}$ and $Q_*=10^{-2}$. \label{fig:ns-r}}
    \vspace{1.2cm}
    \includegraphics{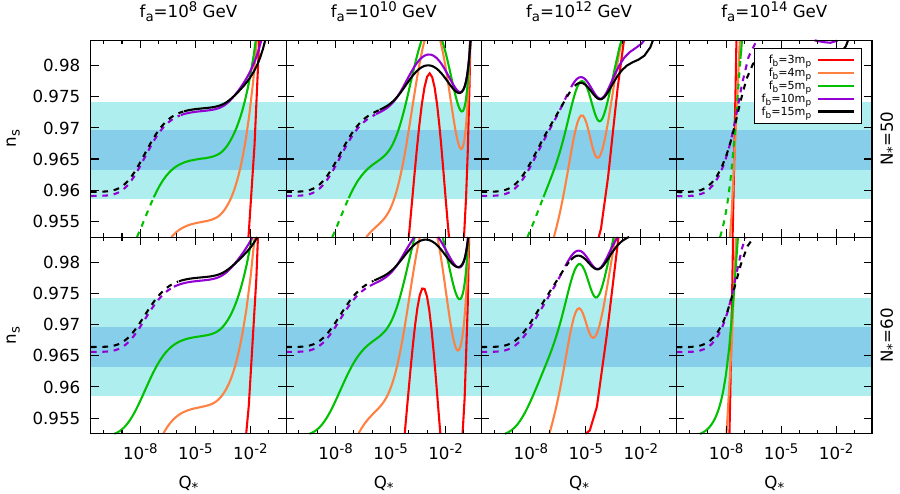}
    \caption{Plot of $n_s$ versus $Q_*$ for different values of $f_a$, $f_b$ and $N_*$. The area shaded in blue marks the observational range for $n_s$ \cite{BK18}, whereas the dashed part of the curves indicates the points where the observational bound for $r$ is not fulfilled, as seen explicitly in Fig.\ \ref{fig:ns-r}.\label{fig:ns-qstar}}
\end{figure}

In the next section we will address the shift symmetric model building constructions but for the moment we will assume an effective Lagrangian containing the terms 
\begin{equation}
	\mathcal{L} \supset \Lambda^4 \left[1-\cos{\left(\frac{\phi}{f_b}\right)}\right] 
	-  \frac{\alpha}{16\pi} \frac{\phi}{f_{a}} F_{\mu\nu} \tilde{F}^{\mu\nu}.
\end{equation}
The new scale opens new windows on the parameter space that can be explored in search of compatibility with the observations. In Fig.\ \ref{fig:ns-r} we show the relation between $n_s$ and $r$ as $Q_*$) ($\alpha$) is varied, whereas in Fig.\ \ref{fig:ns-qstar} we show the results for $n_s$ as a function of $Q_*$. In both figures we compare different values of $f_a$, $f_b$ and $N_*$.



\begin{figure}[ht]
    \centering
    \includegraphics{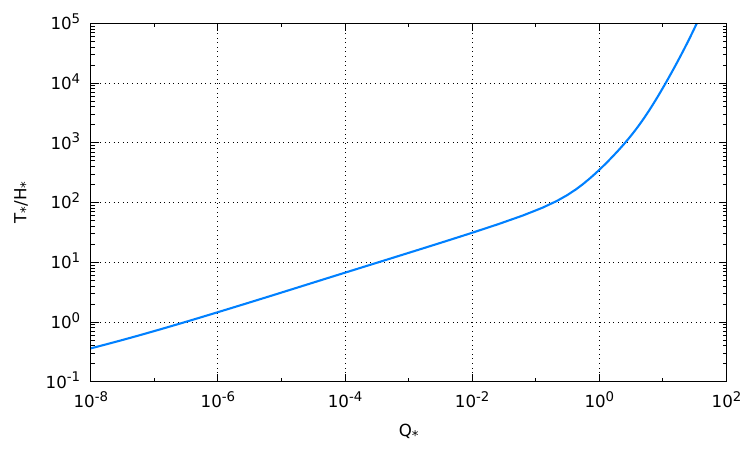}
    \caption{
    Value of $T/H$ versus the dissipative ratio $Q$ at horizon crossing, given by Eq. \eqref{Eq:TH-Q}. 
    \label{fig:TH-Q}}
\end{figure}

In these figures we can observe that introducing the hierarchy yields regions of the parameter space that do match the observations for $n_s$ and $r$. We have found that generally the acceptable values for $Q_*$ in each case, if any, are such that the universe is in the weak or very weak dissipative regime at horizon crossing ($Q_*\ll 1$). { In this regime, dissipation does not impact the background dynamics directly, but observational predictions are already different from those found in CI (as seen in Fig. \ref{fig:fafb_results}). As mentioned before, due to the sphaleron vacuum contribution to the dissipative coefficient one can generate a thermal bath with  $T_*/H_* \sim\mathcal  O(1)$ even for small values of $Q_* \ll 1$. Indeed, combining the expression for the primordial spectrum Eq. \eqref{eq:PRWarm} and the slow-roll Eq. for the radiation energy density:
  \begin{equation}
  \rho_R= g_* \frac{\pi^2}{30} T^4 \simeq \frac{3}{4} Q \dot \phi^2 \,,
  \end{equation}
  one can derive an implicit relation between $Q$ and $T/H$, independently of the potential,
  \begin{equation}
  \left(\frac{T}{H}\right)^4 = 
  \frac{45}{2 \pi^4 g_*P_{\mathcal R}} \left(
  \coth \frac{H}{2T} + \frac{2\sqrt 3 \pi Q}{\sqrt{3+4\pi Q}}\frac{T}{H}\right) G(Q) \,. \label{Eq:TH-Q}
  \end{equation}
This is plotted in Fig. \ref{fig:TH-Q} at horizon crossing, using $P_{\mathcal R}= 2.1 \times 10^{-9}$ \cite{BK18}, where already for $Q_*\simeq 10^{-7}$ we have $T_*/H_* \simeq 1$. Therefore, under the assumption of thermal inflaton fluctuations, this modifies the amplitude of the primordial spectrum in the very WDR,  
\begin{equation}
    P_{\mathcal R} \simeq \left(\frac{H_*}{2\pi}\right)^2\left(\frac{H_*}{\dot \phi_*}\right)^2 \coth\left(\frac{H_*}{2 T_*}\right) \,,
\end{equation}
and therefore both the spectral index and the tensor-to-scalar ratio with respect to the CI predictions. The latter is now suppressed by a factor $1 + 2 n_*$,
\begin{equation}
r \simeq \frac{16 \varepsilon_V^*}{\coth(H_*/(2T_*))}\,,
\end{equation}
and may render the prediction compatible with the observational upper limit on $r$ even for $f_b \sim \mathcal O(10 m_p)$ as seen in Fig. \ref{fig:ns-r}. On the other hand, taking the limit $Q_* \ll 1$ in  the expression for the spectral index \eqref{eq:ns-warm} one has:
\begin{equation}
n_s \sim 1 + 6 \varepsilon_V^* - 2 \eta_V^* + \frac{H_*/T_*}{\sinh(H_*/T_*)} \left( \frac{7 }{4}\varepsilon_V^* - \frac{1}{2}\eta_V^*\right) \,,
\end{equation}
which gives a positive correction to the spectral index when $T_*/H_* \sim\mathcal  O(1)$, shifting the curves in Fig. \ref{fig:fafb_results} to the right.
This shift towards observational allowed values is generally achievable in theory for any value of $f_a$ and $f_b$, but as can be seen in Fig.\ \ref{fig:ns-qstar}, in some cases it requires extremely fine-tuned values of $Q_*$ (i.e., of the coupling $\alpha$), see for example the case $f_a=10^{14}$ GeV in Fig. \ref{fig:ns-qstar}.}
The general behavior seems to be such that as the hierarchy between $f_a$ and $f_b$ increases, so does the range of observationally compatible values for $\alpha$, reducing the need for fine-tuning. In general, a hierarchy of 7 orders of magnitude or higher seems to be necessary to find an acceptable range of CMB-compatible values of $\alpha$.

\begin{figure}[t]
    \centering
    \includegraphics{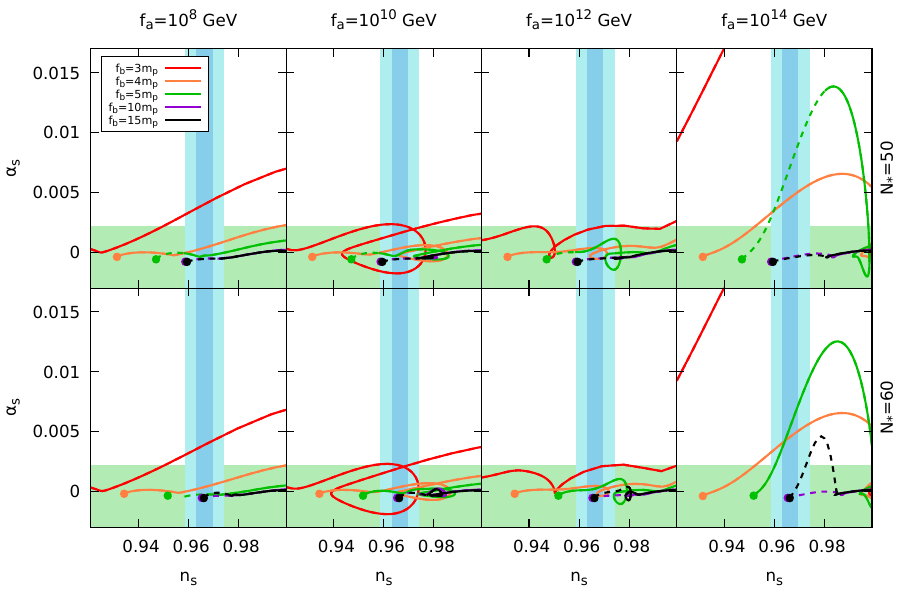}
    \caption{Plot of $\alpha_s$ versus $n_s$ for different values of $f_a$, $f_b$ and $N_*$, with CI values marked as dots in each case. The area shaded in blue marks the observational range for $n_s$ \cite{BK18}, the area shaded in green marks the observational range for $\alpha_s$ \cite{Planck:2018jri} and the dashed lines mark the regimes where the observational bound for $r$ is not fulfilled, as seen in Fig.\ \ref{fig:ns-r}.}
    \label{fig:alphas-ns}
\end{figure}

Moreover, when $n_s$ varies rapidly with $Q_*$, it may indicate a running of the scalar spectral index, $\alpha_s$, out of its observational range. In Fig. \ref{fig:alphas-ns} we show plots of $\alpha_s$ with respect to $n_s$ for all cases studied, marking the observational range for $n_s$, $\alpha_s$ and $r$. Although in figures \ref{fig:ns-r} and \ref{fig:ns-qstar} it may seem that $f_a=10^{14}\,\text{GeV}$ does present observationally-compatible regimes for some values of $f_b$ (save for fine-tuning considerations), Fig. \ref{fig:alphas-ns} shows that this value of $f_a$ is completely excluded since it never fulfills all three observational conditions simultaneously. Because of this, we will no longer include it in the following plots in this paper, and will instead focus on the values of $f_a$ that do present observationally compatible values.

Another case that turns out to be excluded by the running is $f_b=3 m_p$ when $f_a=10^{8}\,\text{GeV}$. In the two remaining columns, $f_a=10^{10}\,\text{GeV}$ and $f_a=10^{12}\,\text{GeV}$, $f_b=3m_p$ behaves as a limiting case, since it does present some values that are compatible with all constraints, but these are found on the edge of the valid range for $\alpha_s$. The only exception is the lower part of the ''loop'' behavior found for $f_b=3m_p$ and $f_a=10^{10}\,\text{GeV}$, which fulfills the condition for $\alpha_s$ in a more reasonable way, but as seen in Fig.\ \ref{fig:ns-qstar} still suffers from fine-tuning issues for the parameter $Q_*$ ($\alpha$).

Finally, it is important to remark that an assumption of a monomial potential $V(\phi)= \lambda m_p^{4-n} \phi^n$ to drive inflation may implicitly break the $f_a=f_b$ condition. The validity of any expansion of a protected periodic potential implies that $\phi_i\ll f_a$. However, the required 50-60 e-folds of accelerated expansion with CMB constraints on the $\lambda$ parameter may set this condition invalid. 
In fact, taking the example of a quadratic potential ($n=2$) with $\alpha=10^{-4}$  and $f_a=10^{8}$ GeV, we find from the  scalar power spectrum, $P_{\mathcal R}$, that  $\lambda\sim 9\times 10^{-15}$, and that at 50 e-folds  before the end of inflation   $\phi_*\sim 2.7 m_p$,   with  $Q_*\sim17 $. This means that, although there is already relevant dissipation, in order to have a period  50-60 efolds of inflation one must break the expansion condition $\phi_i/f_a\sim 7\times 10^{10}\gg1$. 
Alternatively, taking $f_a\ne f_b\sim10\, m_p$ allows for $\phi_i/f_b\sim 0.27$ that may give a more reasonable justification for the expansion of the periodic potential. 

\section{Looking for compatible Model Building }\label{Sec:Modelbulding}

In the previous section we have seen effective models compatible with  observational constrains. The viable parameter space pointed to models where the transition from a weak to strong dissipation regime was happening at CMB scales. 
In scenarios where the hierarchy between $f_a$ and $f_b$ was small, this transition was abrupt and it required an especially tuned effective coupling $\alpha$ to obtained the desirable value of $Q_*$. Moreover, these parameter regions were  excluded by the running of the spectral index $\alpha_s$. 
On the opposite end, for larger hierarchies between the decay constants  a broader range for the effective coupling was allowed, opening a gap for viable model building. In this section we now turn into an analysis on the most common model building options to obtain such hierarchies. Finally, we will also discuss agnostic effective field theory (EFT) bounds on the phenomenological models.

An effective Lagrangian in these models takes the form 
\begin{equation}
	\mathcal{L} \supset \Lambda^{4} \cos\!\left( \frac{\phi}{j f_{a}} \right)
	+ k \frac{\bar{\alpha}}{8\pi} \frac{\phi}{f_{a}} F_{\mu\nu} \tilde{F}^{\mu\nu},
\end{equation}
where the first term, as before, is the effective potential of the axion generated, for instance, by non perturbative dynamics of some confining gauge group with a confinement scale $\Lambda$. The effective field range of the axion is then $f_b=j f_a$. Here $\bar{\alpha}=g^2/(4\pi)$ and $g$ is the gauge coupling. 
The coefficient $k$ represents an enhanced coupling between the axion field and the gauge fields, and  must be an integer due to the axion shift symmetry. In the previous section we took $k=1$. 

It is possible to enhance separately the parameters $k$  and $j$. Solutions like the clockwork models \cite{Choi:2015fiu,Kaplan:2015fuy}, and alignment conditions \cite{Kim:2004rp} can increase the field range $f_b=j f_a$. On the other hand, the inclusion of matter fields such as vector-like fermions with large charges under the gauge group or large PQ charge, and kinetic mixing between multiple axions \cite{Babu:1994id} may increase $k$. 

These specific mechanisms have different theoretical constrains and present limitations for generating arbitrarily large couplings and hierarchies. In particular, see \cite{Agrawal:2018mkd,Bagherian:2022mau} for more details, clockwork models are limited to an enhancement on the effective field range of $j\lesssim f_b/\Lambda$ that is equivalent to the condition $\Lambda\lesssim f_a$. On the other hand, perturbativity constrains bound the factor $k$ to $k\lesssim\bar{\alpha}^{-1}$. With the parametrization discussed in the previous section ($k=1$) it  implies that $\alpha= \bar{\alpha}\lesssim1$.

On more general grounds, the still viable phenomenological models can also be challenged in a  model independent description through EFT considerations and partial wave unitarity bounds.
The basic idea of such bounds, derived in this context for a $SU(2)$ gauge group in \cite{Bagherian:2022mau}, is to use the unitarity of the $S$-matrix together with a decomposition of scattering states into angular momentum modes to bound partial-wave amplitudes, which are functions of center-of-mass energy alone. By studying the scattering of gauge fields and axions mediated by the $\phi F \tilde{F}$ coupling one finds 
\begin{equation}
    \Lambda\lesssim 32  \pi^{3/2} \frac{f_a}{\alpha}\label{eq:unitarity},
\end{equation}
which is naturally less restrictive than the clockwork condition $\Lambda\lesssim f_a$ mentioned above.

\begin{figure}[t]
    \centering
    \includegraphics{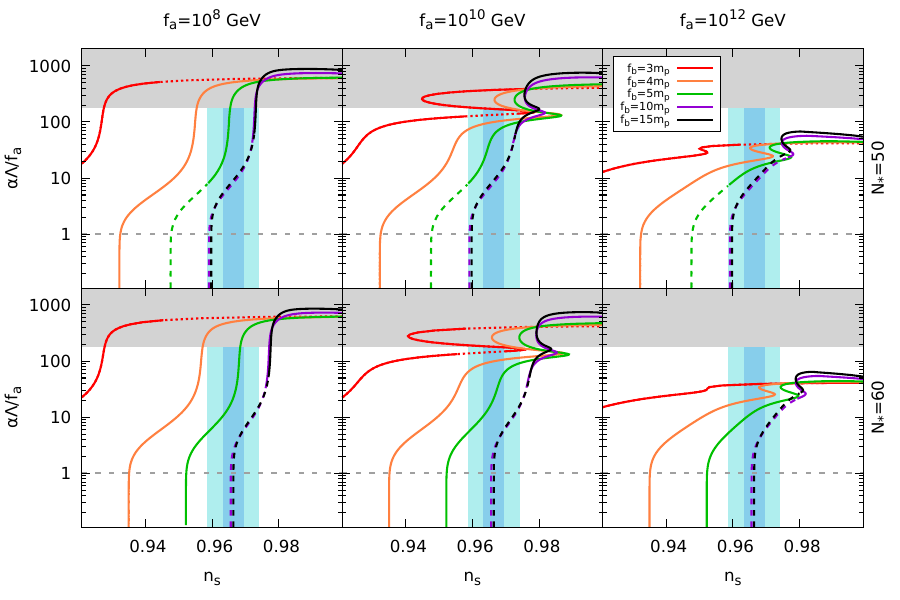}
    \caption{Plot of $\alpha\Lambda/f_a$ versus $n_s$. Here, the shaded gray area marks the values ruled out by the unitarity constraint (\ref{eq:unitarity}), while the blue area marks the observational range for $n_s$ \cite{BK18}. The dashed part of the curves mark the regions where $r$ is outside its observational range as seen in Fig.\ \ref{fig:ns-r}, and the dotted ones mark the analogous for $\alpha_s$ as seen in Fig.\ \ref{fig:alphas-ns}. The gray dashed line at $\alpha \Lambda/f_a=1$ shows that the clockwork condition is never fulfilled within observational range, since $\alpha\Lambda/f_a>1$ implies $\Lambda/f_a>1$.}
    \label{fig:alphalambdafa-ns}
\end{figure}

In Fig.\ \ref{fig:alphalambdafa-ns} we contrast the model building bounds with the observational constrains. Immediately, since there are no observationally valid regions below the gray dashed line $\alpha \Lambda/f_a=1$, one can see that the clockwork condition is never fulfilled. Regarding  the unitarity bound introduced in (\ref{eq:unitarity}), restricted within the gray shaded area,  we see that for larger hierarchies between $f_a$ and $f_b$, some of the observation-compatible values studied before are also ruled out by this condition, the most notable being the case with $f_a=10^8\,\text{GeV}$ and $f_b=4m_p$ which is now completely excluded. The $f_a=10^{10}\,\text{GeV}$ cases see their available parameter space reduced, but still present a range that is compatible with the observations, while the $f_a=10^{12}\,\text{GeV}$ cases are unaffected by the unitarity constraint.
We can conclude that there are parameters for which the two-scale model is phenomenologically viable, but the clockwork mechanism is not a viable explanation for the hierarchy.

\section{Discussion: The possible MWI model}
\label{sec:Disc}

To conclude our analysis, we would like to recover the discussion we omitted at the end of Section \ref{sec:section2} regarding the inflaton oscillation frequency $\omega$. To ensure that a thermal bath can be generated before CMB scales,  we have considered ${\omega=m}$ throughout this work.  In the literature, the possibility of taking $\omega^2=\max(0,\partial_\phi^2V)$  has also been proposed \cite{Klose:2022rxh}. We would like to state that contrary to what is stated in that reference, for harmonic potentials, this choice does affect both the qualitative and quantitative evolution of the model.

Typically in cold initial conditions ($T_i\sim 0$) with harmonic potentials, equation \eqref{eq:potential1}, one must generally start inflation near the top of the potential ($\phi\sim \pi f_b$), where the second derivative is negative, in order to obtain 50-60 e-folds of inflation\footnote{Although this is not necessary for larger scales $f_b\gtrsim 10\,m_p$, such values of $f_b$ are much less relevant for the observations than those around $m_p \lesssim f_b\lesssim 10\,m_p$, where the whole height of the potential is necessary in order to obtain 50 e-folds of inflation. \label{fn:SDR}}.  We have seen that,  in order to satisfy observations, when taking $\omega=m$, { a thermal bath should  already be present at the time of horizon crossing.}

\begin{figure}[t]
    \centering
    \includegraphics{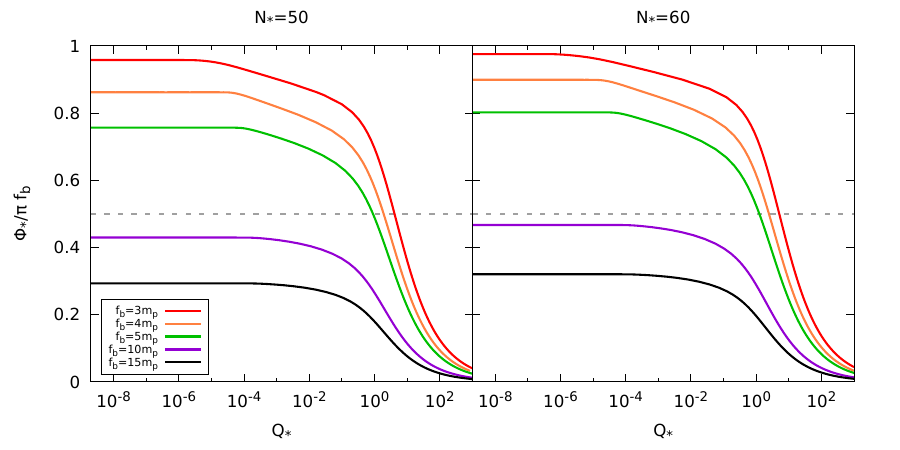}
    \caption{Value of the inflaton at horizon crossing, $\phi_*$ normalized by $\pi f_b$, with respect to the dissipative ratio at horizon crossing, $Q_*$. The data in this plot corresponds to $f_a=10^{12}\,\text{GeV}$, but other values of $f_a$ present the same behavior, up to very small variations. The dashed gray line marks the value $\phi_*/f_b=\pi/2$.}
    \label{fig:phistar-qstar}
\end{figure}

\begin{figure}[t]
    \centering
    \includegraphics{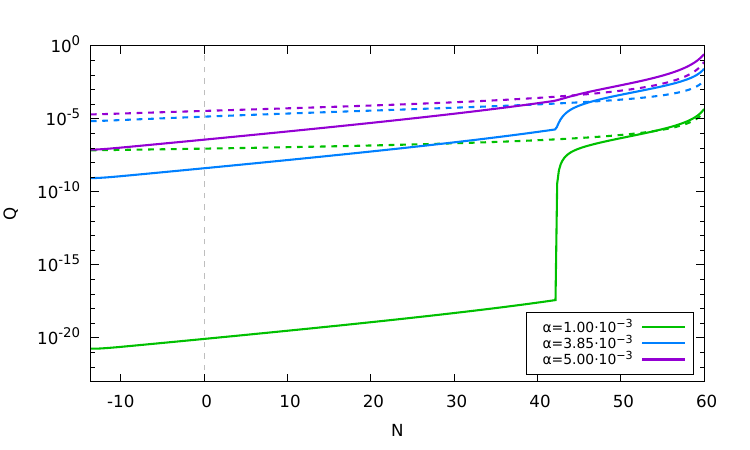}
    \caption{Evolution of the dissipative ratio $Q$ with respect to the number of e-folds left until the end of inflation (with $N=0$ marking horizon crossing), for the parameters $f_a=10^{12}\,\text{GeV}$, $f_b=5m_p$ and $N_*=60$, using different values of $\alpha$. We compare the results for $\omega^2=\max(0,\partial_\phi^2V)$ (solid lines) with the setup $\omega=m$ used in previous sections (dashed lines). Observational predictions and values of $Q_*$ for each case are included in Table \ref{tab:QComparison}.}
    \label{fig:Q-comparison}
\end{figure}

Even in this case, the possible MWI models still require $\phi_*>\pi f_b/ 2$, see Fig.\ \ref{fig:phistar-qstar}, and the evolution of $\omega$ may have an impact of the model evolution. On  closer inspection (see the dashed lines in Fig.\ \ref{fig:Q-comparison}), still considering $\omega=m$, { although the cubic dissipative coefficient will dominate by the end of inflation, we need the $\omega$ contribution at horizon crossing that ensures $Q_* \simeq \mathcal O(10^{-7}-10^{-5})$, enough to have $T_*/H_* \sim \mathcal O(1)$, and the good behavior required to satisfy the running of the spectral index constrains.} In contrast, the solid lines in  Fig. \ref{fig:Q-comparison} represent the evolution with $\omega^2=\max(0,\partial_\phi^2V)$, { with a visible transition roughly 20 e-folds before the end of inflation, when the inflaton start evolving in the positive curvature part of the potential, and $\omega$ contributions to dissipation help increasing the thermal bath.} In Table \ref{tab:QComparison} we present the predictions for the parameters in Fig. \ref{fig:Q-comparison}. Among these examples, $\alpha=10^{-3}$ is compatible with all observational bounds for $\omega=m$, $\alpha=3.85\cdot 10^{-3}$ is compatible for $\omega^2=\max(0,\partial_\phi^2V)$, and $\alpha=5\cdot 10^{-3}$ is excluded in either case\footnote{However, for these $\alpha$ parameters, when considering both Planck-BK18 and ACT bounds $\omega=m$ is always compatible with observations.  On the other hand, $\omega^2=\max(0,\partial_\phi^2V)$ is never compatible with ACT observations.}.
This states the impact of the vacuum oscillations even in the weak dissipation regime where a non zero temperature is already present.

\begin{table}[t]
\centering %
\begin{tabular}{|c|c|c|c|c|c|c|}
\cline{2-7}
\multicolumn{1}{c|}{ } & \multicolumn{3}{c|}{$\omega=m$} & \multicolumn{3}{c|}{$\omega=\max{(0,\partial_{\phi}^{2}V)}$}\tabularnewline
\hline 
$\alpha$ & $n_{s}$ & $r$ & $Q_{*}$ & $n_{s}$ & $r$ & $Q_{*}$\tabularnewline
\hline 
$1.00\cdot10^{-3}$  & $0.965$ & $2.02\cdot10^{-2}$ & $9.17\cdot10^{-8}$ & $\mathbf{0.952}$ & $3.26\cdot10^{-2}$ & $8.58\cdot10^{-21}$\tabularnewline
\hline 
$3.85\cdot10^{-3}$  & $\mathbf{0.977}$ & $4.62\cdot10^{-3}$ & $1.42\cdot10^{-5}$ & $0.961$ & $3.07\cdot10^{-2}$ & $4.17\cdot10^{-9}$\tabularnewline
\hline 
$5.00\cdot10^{-3}$  & $\mathbf{0.974}$ & $3.44\cdot10^{-3}$ & $3.50\cdot10^{-5}$ & $\mathbf{0.990}$ & $1.45\cdot10^{-2}$ & $3.77\cdot10^{-7}$\tabularnewline
\hline 
\end{tabular}\caption{Observational predictions and values of $Q_{*}$ for the different
cases presented in Fig.\ \ref{fig:Q-comparison}. These results were
obtained using the parameters $f_{a}=10^{12}\,\text{GeV}$, $f_{b}=5m_{p}$
and $N_{*}=60$. Values of $n_s$ highlighted in bold are excluded by observational constraints \cite{BK18}.}\label{tab:QComparison}
\end{table}

\begin{figure}[t]
    \centering
    \includegraphics{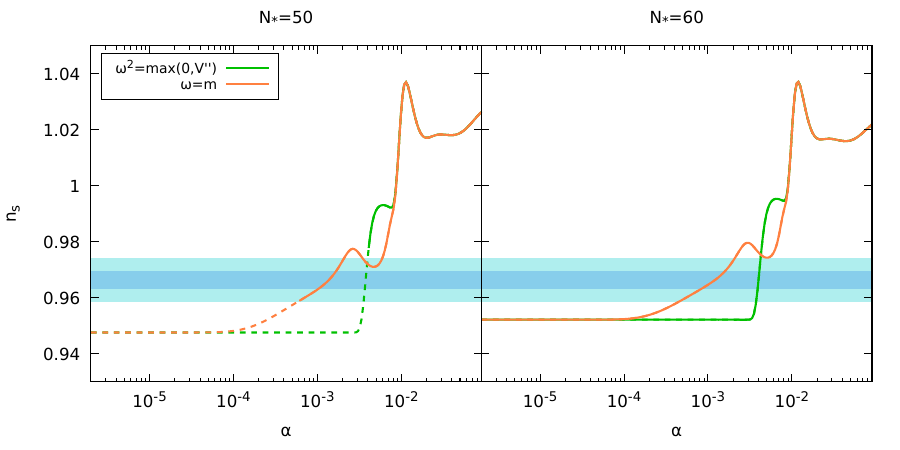}
    \caption{Results of $n_s$ with respect to $\alpha$ for the parameters $f_a=10^{12}\,\text{GeV}$, $f_b=5m_p$, comparing the choices of $\omega^2=\max(0,\partial_\phi^2V)$ and $\omega=m$. As in previous plots, the area shaded in blue marks the observational range for $n_s$ \cite{BK18}, and the dashed part of the curves marks the regime where the observational bound for the tensor-to-scalar ratio is not fulfilled as seen in Fig.\ \ref{fig:ns-r}.}
    \label{fig:ns-alpha-variable}
\end{figure}

Focusing on $\omega^2=\max(0,\partial_\phi^2V)$, we find that when $|\phi_*| > \frac{\pi}{2} f_b$, a non-zero initial temperature must be imposed to avoid the cold inflation CMB constraints. For $|\phi_*| \leq \frac{\pi}{2} f_b$, a thermal bath may instead be dynamically generated through the evolution of $\omega$. However, aside from specific configurations already excluded by the tensor-to-scalar ratio (see footnote~\ref{fn:SDR}), { this mechanism with $\omega^2 >0$ already at horizon-crossing would be viable only in the strong dissipative regime,  which is itself incompatible with the observed value of $n_s$.
  Fig.\ \ref{fig:ns-alpha-variable} shows the evolution of the spectral index as the coupling constant $\alpha$ (and with it $Q_*$) is varied, compared to the reference case $\omega = m$. One can see the abrupt transition when $\alpha$ is varied from the case with a negligible thermal bath all the way along inflation (recovering the predictions for CI), to have those effects affecting the evolution and the predictions.} 
%

In summary,  we verify that the precise definition of $\omega$  plays a decisive role in the phenomenology of the model and in its observational predictions, making it a central 
aspect to be further investigated in future work.

\section{Conclusions} \label{sec:Concl}

In this paper we have  studied the observationally viable Minimal Warm Inflation models with periodic potentials. We  started by revising  the simplest scenario with the same decay constant in the axion potential and in the dissipation coefficient ($f_a=f_b$). In this case, neither in the strong dissipation regime, where a periodic potential had been shown to be incompatible with a period of 50-60 efolds of inflation \cite{Zell:2024vfn}, nor in the weak dissipation regime, the observational limits at CMB scales can be met. 

If a hierarchy between the decay constants is allowed, the possible parameter space opens, and models with $4\, m_p\lesssim f_b\lesssim5\, m_p$ and $10^{8}\, \mathrm{GeV}\lesssim f_a\lesssim10^{12}\,\mathrm{GeV}$ are then phenomenologically viable. Meaning, a hierarchy of $f_b/f_a\sim O(10^7 -10^{11})$ is required to match CMB constrains. 
Moreover, as stated at the end of Section \ref{Sec:parameterspace}, polynomial potentials,  taken as expansions of periodic potentials, also require these hierarchies.

However, when looking for model building possibilities, we have verified that the popular clockwork mechanism cannot generate such hierarchies.  
Nevertheless, when studying partial wave unitary bounds we verify that EFTs with  more restricted hierarchies would represent valid constructions. 
To our knowledge, there are no identified models that can explain the needed hierarchy, and creativity is required to go beyond the standard clockwork mechanism.

To conclude, we note that in these models several questions remain open. These could  affect the model evolution and observational viability. Examples are the thermalization of the inflaton and the ambiguous physical interpretation of the frequency $\omega$ for more complex potentials with negative curvatures. 
Another effect that may have significant implications for parametric model-building is the recently reported emergence of an effective chemical potential. As discussed in \cite{Broadberry:2025ggb}, this effect may suppress thermal friction. However, quantifying its impact on the non-perturbative dynamics is highly non-trivial and requires further dedicated study. Nevertheless, a naive analysis suggests that its primary influence would be on the dynamics themselves rather than on the resulting power spectrum. Since most of our results lie within the WDR, we do not expect major qualitative modifications, but rather shifts in the viable parameter space.
Addressing these issues, together with the associated model-building challenges, may open new avenues for investigation and provide further insight into the viability of these scenarios.

\acknowledgments

MBG  and ATM acknowledge partial support by PID2022-140831NB-I00 funded by MICIU/AEI/10.13039/501100011033 and FEDER,UE. This work was supported by national funds by FCT - Fundação para a Ciência e Tecnologia, I.P., through the research projects with DOI identifiers 10.54499/UID/04564/2025 and by the project 10.54499/2024.00252.CERN funded by measure RE-C06-i06.m02 – “Reinforcement of funding for International Partnerships in Science, Technology and Innovation” of the Recovery and Resilience Plan - RRP, within the framework of the financing contract signed between the Recover Portugal Mission Structure (EMRP) and the Foundation for Science and Technology I.P. (FCT), as an intermediate beneficiary.
\appendix


\section{Scalar spectral index in Warm Inflation}
\label{sec:ns-warm}
As discussed in section \ref{sec:section2}, the expression for the scalar amplitude of CMB perturbations in Warm Inflation is given by  \cite{Bastero-Gil:2018uep,Spectrum1,Spectrum2,Spectrum3}:

\begin{equation}
    P_{\mathcal R} = \left(\frac{H_*}{2\pi}\right)^2\left(\frac{H_*}{\dot \phi_*}\right)^2\left(\coth\left({\frac{H_*}{2T_*}}\right) + \frac{2\sqrt 3 \pi Q_*}{\sqrt{3+4\pi Q_*}}\frac{T_*}{H_*}\right) G(Q_*),
    \label{eq:PRWarmAppendix}
\end{equation}
where we have assumed that the inflaton's fluctuations are thermal and thus $n_*\approx(e^\frac{H_*}{T_*}-1)^{-1}$. In order to obtain an expression for the spectral index $n_s$, one must derive the previous expression with respect to the comoving scale $k$ at horizon crossing, which is approximately equivalent to deriving with respect to the number of e-folds,
\begin{equation}
    n_s-1\equiv \frac{d\ln P_{\mathcal R}}{d\ln k}\approx\frac{d\ln P_{\mathcal R}}{dN}.
\end{equation}
Deriving equation (\ref{eq:PRWarmAppendix}) gives the following result \cite{Das:2022ubr}:
\begin{multline}
    n_s-1\approx
    \frac{2\eta_V^*}{1+Q_*}
    -6\epsilon_H^*
    +\frac{2Q_*}{1+Q_*}\left.\frac{d\ln Q}{dN}\right|_*
    +\frac{G'(Q_*)}{G(Q_*)}\left.\frac{dQ}{dN}\right|_*
    \\
    +\frac{1}{B_Q} \left( A_Q \left(\frac{3+2 \pi  Q_*}{3+4 \pi  Q_*} \right)\left.\frac{d\ln Q}{dN}\right|_*+\left( \frac{H_*/T_*}{2\sinh^2{ (H_*/2T_*)}}+A_Q\right)\left.\frac{d\ln (T/H)}{dN}\right|_*\right), \label{eq:ns-warm}
\end{multline}
where for ease of notation we have defined
\begin{equation}
    A_Q\equiv\frac{2 \sqrt{3}\pi  Q_*}{\sqrt{3+4 \pi  Q_*}}\frac{T_*}{H_*},\,\,\,\,\,\,\,\,\,\,B_Q=\left(\coth\left({\frac{H_*}{2T_*}}\right) +A_Q\right).\label{eq:appendix-AQ}
\end{equation}
The expression for $n_s$ depends on the derivatives of $Q$ and $T/H$ with respect to the number of e-folds. By deriving the slow-roll equations and the definition of $Q$, we arrive at the following expressions for these derivatives,
\begin{equation}
    \frac{d\ln\phi}{dN} = -\frac{\sigma_V}{1+Q},
\end{equation}
\begin{equation}
    \frac{d\ln Q}{dN}
    =
    q_T \frac{d\ln T}{dN} +p_\omega \frac{d\ln \omega}{dN}  +\varepsilon_H,\label{eq:appendix-dlnQ}
\end{equation}
\begin{equation}
    \frac{d \ln (T/H)}{dN}=\frac{d \ln T}{dN}+\varepsilon_H=\frac{(1-Q)\left(
    \varepsilon_H+p_\omega\frac{d\ln\omega}{dN}
    \right)  -2(\eta_V- \varepsilon_V)}{4(1+Q)-q_T(1-Q)}+\varepsilon_H,\label{eq:appendix-dlnTH}
\end{equation}
where we have defined the following parameters,
\begin{equation}
     \sigma_V\equiv \frac{m_p^2}{\phi^2}\frac{\partial\ln V}{\partial\ln \phi},\,\,\,\,\,\,\,\,p_\omega\equiv \frac{\partial \ln \Upsilon}{\partial \ln \omega},\,\,\,\,\,\,\,\,  q_T\equiv \frac{\partial \ln \Upsilon}{\partial \ln T}.
\end{equation}
We have included the dependence of $\Upsilon$ on $\omega$ as a variable, so that all expressions are valid when a dynamical expression for $\omega$ is used, such as $\omega^2=\max(0,\partial_\phi^2V)$. Its derivative with respect to $N$ can be written as
\begin{equation}
    \frac{d\ln\omega}{dN}=\frac{d\ln\omega}{d\phi}\frac{d\phi}{dN}= -\frac{d\ln\omega}{d\ln \phi}\frac{\sigma_V}{(1+Q)}.
\end{equation}
In the simplified case $\omega=m$ this is equal to zero, and only the $T$-dependence of $\Upsilon$ remains relevant for the derivatives.

\section{Non-thermal inflaton fluctuations}\label{Sec:thermalinf}

In section \ref{sec:WICMB} we discussed the choice for the distribution of the inflaton's fluctuations $n_*$ at the point of horizon crossing, which appears in equation (\ref{eq:PRWarm}). Depending on the strength of the interactions between the inflaton's fluctuations and the radiation bath, it can be assumed to take a certain value between the Bunch-Davies vacuum and a Bose-Einstein distribution,
\begin{equation}
    0\leq n_*\lesssim (e^{H_*/T_*}-1)^{-1}.
\end{equation}
The first equality holds if the inflaton's fluctuations are completely decoupled from the thermal bath, and the second one holds if they are completely thermalized \cite{Bastero-Gil:2018uep}. In this work, for simplicity and to compare with previous studies we have studied the latter case, assuming the inflaton's fluctuations are thermal. However, it is also possible to assume the opposite and take $n_*=0$, resulting in different results for primordial spectrum predictions. In that case, we find the following expressions for $P_\mathcal R$ and $n_s$:
\begin{equation}
    P_{\mathcal R} = \left(\frac{H_*}{2\pi}\right)^2\left(\frac{H_*}{\dot \phi_*}\right)^2\left(1 + \frac{2\sqrt 3 \pi Q_*}{\sqrt{3+4\pi Q_*}}\frac{T_*}{H_*}\right) G(Q_*),
\end{equation}
\begin{multline}
    n_s-1\approx\frac{2\eta_V^*}{1+Q_*} -6\epsilon_H^* +\frac{2Q_*}{1+Q_*}\left.\frac{d\ln Q}{dN}\right|_* +\frac{G'(Q_*)}{G(Q_*)}\left.\frac{d\ln Q}{dN}\right|_*  \\+\frac{ A_Q }{1+A_Q} \left( \left(\frac{3+2 \pi  Q}{3+4 \pi  Q} \right)\left.\frac{d\ln Q}{dN}\right|_*+\left.\frac{d\ln (T/H)}{dN}\right|_*\right),
\end{multline}
where $A_Q$ is defined in equation (\ref{eq:appendix-AQ}), and in this case $B_Q=1+A_Q$. The derivatives of $Q$ and $T/H$ can be found in (\ref{eq:appendix-dlnQ}) and (\ref{eq:appendix-dlnTH}), respectively.

\begin{figure}[t]
    \centering
    \includegraphics{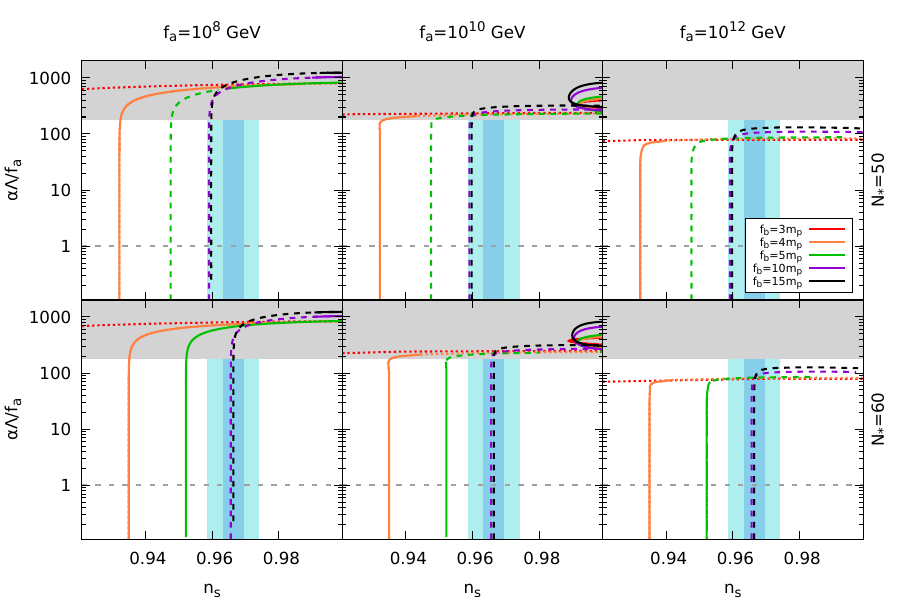}
    \caption{Plot of $\alpha\Lambda/f_a$ versus $n_s$, assuming completely non-thermal inflaton fluctuations ($n_*=0$). Observational and model-building constraints are represented in the same manner as in Fig.\ \ref{fig:alphalambdafa-ns}.}
    \label{fig:nonthermal}
\end{figure}

Although the choice of $n_*$ does not directly impact the evolution of the system, its absence results in different predictions for CMB parameters. To check the predictions of this setup, we repeat the analysis seen in Fig.\ \ref{fig:alphalambdafa-ns}, obtaining the results represented in Fig.\ \ref{fig:nonthermal}. There, one can see that, for the intermediate case $f_a=10^{10}\,\text{GeV}$, neither the unitarity bound (\ref{eq:unitarity}) or the CMB constraints for $n_s$, $r$ and $\alpha_s$ are fulfilled for any value of $\alpha$ or $f_b$. The $f_a=10^{12}\,\text{GeV}$ case shows that increasing $f_a$ allows the system to fulfill the unitarity bound, but it is still ruled out by CMB constraints. On the other side, in the $f_a=10^{8}\,\text{GeV}$ case we see that decreasing $f_a$ does yield some CMB-compatible results, but these are still excluded by the unitarity bound.
In other words, we have observed that the fully non-thermal inflaton scenario requires a minimum hierarchy of $f_b/f_a\gtrsim O(10^{11})$ in order to fulfill CMB constraints, much larger than the thermalized inflaton case (around $f_b/f_a\gtrsim O(10^7)$, as seen in section \ref{Sec:parameterspace}). It is impossible to achieve such a hierarchy without breaking the unitarity bound introduced in section \ref{Sec:Modelbulding}, therefore we may conclude that assuring that the inflaton's fluctuations are thermalized is a requirement for the model to be viable.



\newpage
\bibliographystyle{JHEP}
\bibliography{mybibliography}

\end{document}